\newcommand{\tr}{\rm tr \,}
\documentclass[prd,twocolumn,showpacs,preprintnumbers,superscriptaddress]{revtex4}
\usepackage{graphicx}% Include figure files
\usepackage{dcolumn}% Align table columns on decimal point
\usepackage{bm}% bold math

\begin{document}

%\preprint{APS/123-QED}

\title{Chiral dynamics for exotic open-charm resonances}% Force line breaks with \\
\author{M.F.M. Lutz}
 \affiliation{Gesellschaft f\"ur Schwerionenforschung (GSI)\\
Planck Str. 1, 64291 Darmstadt, Germany }
\author{E.E. Kolomeitsev}%
\affiliation{%
The Niels Bohr Institute, Blegdamsvej 17, DK-2100 Copenhagen,
Denmark
}%


\begin{abstract}
We review the open-charm spectrum of chiral excitations in QCD. At
leading order in a chiral expansion a parameter-free prediction is
obtained for the coupled-channel scattering of Goldstone bosons
off open-charm mesons and baryons with $0^-,1^-$ and
$\frac{1}{2}^+$ quantum numbers. The recently announced narrow
$D_s(2317)$ and $D_s(2463)$ mesons observed by the BABAR and CLEO
collaborations are reproduced. Also the baryon states
$\Lambda_{c}(2593), \Lambda_{c}(2880)$ and $\Xi_{c}(2790)$
discovered earlier by the CLEO collaboration are recovered
naturally. Besides describing resonances with conventional quantum
numbers additional multiplets with members that carry exotic
quantum numbers are predicted. In particular, so far unobserved
narrow isospin-singlet open-charm mesons with negative strangeness
are generated dynamically. Similarly, narrow isospin-doublet
open-charm baryons with positive strangeness are suggested.
\end{abstract}


\maketitle

\section{Introduction}

\vskip-0.2cm Recently a new narrow state of mass 2.317 GeV that
decays strongly into $D^+_s\,\pi^0$ was announced \cite{BaBar}.
This result was confirmed \cite{CLEO} and a second narrow state of
mass 2.463 GeV decaying into $D_s^* \pi^0$ was observed. Such
states were first predicted in \cite{BH94,NRZ93} based on the
chiral quark model \cite{BH94,NRZ93,BEH03,NRZ03}.  The latter
implies the heavy-light $0^+,1^+$ resonance states to form an
anti-triplet representation of the SU(3) group. If one insists on
a non-linear realization of the chiral SU(3) group only, excluding
any further model assumptions, an a priori statement can not be
made for the existence of chiral partners of any given state. Thus
it is it is important to study the heavy-light meson and baryon
resonances in great detail \cite{all-papers}

\vskip+0.04cm In this talk we review a particularly important
aspect of open-charm resonances: the role of coupled-channel
effects. It is long known that light scalar mesons can be
described quantitatively in terms of coupled-channel dynamics
without assuming the existence of light quark-antiquark states
with scalar quantum numbers
\cite{gen-scalars}. %\cite{Rupp86,WI90,JPHS95,OOP99,NVA02,NP02}.
More recently it was pointed out by the authors \cite{LK03} that
similarly the light axial-vector meson spectrum of QCD is
naturally explained in terms of chiral coupled-channel dynamics.
The leading term of the chiral Lagrangian written down for the
interaction of Goldstone bosons with light vector mesons generates
the axial-vector spectrum dynamically. Moreover, chiral dynamics
has the potential to predict the existence of exotic SU(3)
multiplets. For instance in \cite{Copenhagen} weak attraction for
s-wave scattering of Goldstone bosons off the decuplet ground
states was found in the 27-plet channels. This may lead to the
existence of penta-quark type resonance states in the $K
\Delta$-channel. Since chiral coupled-channel dynamics is able to
predict the s- and d-wave baryon resonance spectrum in terms of
the leading order interaction term \cite{Granada,Copenhagen} it is
natural to expect coupled-channel dynamics to play also a crucial
role in the physics of open-charm meson and baryon resonances. The
possible importance of coupled-channel effects for the heavy-light
meson resonances was anticipated by Van Beveren and Rupp
\cite{BR03}. Systematic computations carried out by the authors
\cite{KL04,LK04,HL04} based on the chiral Lagrangian  will be
discussed here. For the possible role of diquark correlations in
scalar open-charm meson resonances we refer to the talk of
Terasaki \cite{Terasaki,Terasaki-talk}.

\vskip-0.04cm A coupled-channel effective field theory for the
scattering of Goldstone bosons off heavy-light meson and baryon
fields is presented \cite{LK00,LK01,LH02,LK02,Granada,Copenhagen}.
It relies on the chiral SU(3) Lagrangian with heavy-light
$J^P\!=\!0^-$, $J^P\!=\!1^-$ and $J^P\!=\!\frac{1}{2}^+$ fields,
that transform non-linearly under the chiral SU(3) group. The
major result in the open-charm meson sector is the prediction that
there exist chiral excitations in QCD with scalar and axial-vector
quantum numbers forming anti-triplet and sextet representations of
the SU(3) group. This differs from the results of the chiral quark
model leading to anti-triplet states only \cite{NRZ03}. It is
pointed out that the heavy-quark symmetry, which suggests a
degeneracy of scalar and axial-vector resonances at leading order,
is naturally recovered in chiral coupled-channel theory. Further
results are derived for open-charm baryon states. The s-wave
scattering of Goldstone bosons off the anti-triplet and sextet
ground states generates dynamically one anti-triplet, two sextet
and one anti-quindecimet of states. This implies the existence of
exotic penta-quark-like states with open-charm quantum numbers.

\vskip-0.04cm Particular results concern the 'heavy' SU(3) limit
with $m_{\pi ,\eta, K} \simeq 500$ MeV in which we predict bound
states rather than resonance states typically. This observation
should facilitate the verification of chiral coupled-channel
dynamics via un-quenched QCD lattice simulations. In the 'light'
SU(3) limit with $m_{\pi ,\eta, K} \simeq 140$ MeV no clear
resonance nor bound-state signals persist. The latter prediction
poses a true challenge for un-quenched QCD lattice simulation,
requiring small current quark masses as well as the analysis of
extremely broad spectral distributions from Euclidian-space
resonance propagators.

\begin{widetext}

\begin{figure*}[t]
\begin{center}
\includegraphics[angle=-90,width=18.0cm,clip=true]{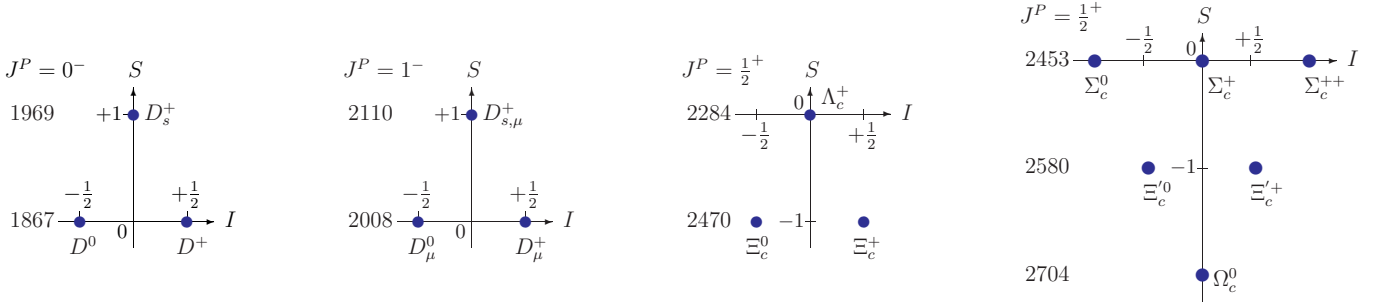}
\end{center}
\caption{Isospin ($I$) and strangeness ($S$) quantum numbers of
open-charm mesons and baryons ground states with $J^P=0^-,1^-,
\frac{1}{2}^+$. The isospin averaged masses in units of MeV of the
various states are indicated. \label{fig:0}}
\end{figure*}

\section{Chiral coupled-channel dynamics: the $\chi$-BS(3) approach}

The starting point of our study is the chiral SU(3) Lagrangian. We
identify the leading-order term \cite{Wein-Tomo,Wise92,YCCLLY92,BD92} describing the interaction
of Goldstone bosons with open-charm pseudo-scalar, vector mesons and
spin-$1/2$ baryons,
\begin{eqnarray}
{\mathcal L}(x) &=& \frac{1}{8\,f^2}\,{\tr }\,\big( \left(  \big[
P_{\phantom{}}(x)\, (\partial^\nu P^\dagger_{\phantom{}}(x) )
-(\partial^\nu P_{\phantom{}}(x))\,P^\dagger_{\phantom{}}(x) \big]
- \big[ P^\mu(x)\, (\partial^\nu P^\dagger_\mu(x)
)-(\partial^\nu P^\mu(x))\,P^\dagger_\mu(x) \big]
\right)\,\big[\Phi (x) , (\partial_\nu\,\Phi(x))\big]_- \big)
\nonumber\\
&+& \frac{i}{16\,f^2}\,{\tr }  \big( \bar H_{[\bar
3]}(x)\,\gamma^\mu\, \big[ H_{[\bar 3]}(x) \,,[\Phi (x) ,
(\partial_\mu\,\Phi(x))]_-\big]_+ \big) +\frac{i}{16\,f^2}\,\tr
\big( \bar H_{[6]}(x)\,\gamma^\mu\, \big[ H_{[6]}(x) \,,[\Phi (x)
, (\partial_\mu\,\Phi(x))]_-\big]_+ \big)  \,. \label{WT-term}
\end{eqnarray}
\end{widetext}

Here $\Phi$ is the Goldstone bosons field, $P$ and $P_\mu$ are the
massive pseudo-scalar and vector-meson fields and and $H_{[\bar
3]}$ and $H_{[6]}$ the massive baryon fields. The $J^P=0^-$ and
$J^P=1^-$ fields, $P$ and $P_\mu$, form anti-triplet
representations of the SU(3) flavour group. The
$J^P=\frac{1}{2}^+$ baryon fields $H_{[\bar 3]}$ and $H_{[6]}$
transform as anti-triplet and sextet. The corresponding isospin
(I) and strangeness (S) quantum numbers are recalled in Fig.
\ref{fig:0}. Finally the parameter $f$ in (\ref{WT-term}) is known
from the weak decay process of the pions. We use $f= 90$ MeV.

The scattering problem decouples into seven orthogonal channels
specified by isospin and strangeness quantum numbers.
Heavy-light meson and baryon resonances with quantum numbers $J^P\!=\!0^+$,
$J^P\!=\!1^+$ and $J^P\!=\!\frac{1}{2}^-$ manifest themselves as poles in the s-wave
scattering amplitudes, $M^{(I,S)}_{J^P}(\sqrt{s}\,)$, which in the
$\chi-$BS(3) approach \cite{LK02,LK03} take the simple form
\begin{widetext}
\begin{eqnarray}
&&  M^{(I,S)}_{J^P}(\sqrt{s}\,) = \Big[ 1-
V^{(I,S)}_{J^P}(\sqrt{s}\,)\,J^{(I,S)}_{J^P}(\sqrt{s}\,)\Big]^{-1}\,
V^{(I,S)}_{J^P}(\sqrt{s}\,)\,,\qquad
V^{(I,S)}_{\frac{1}{2}^+}(\sqrt{s}\,) =
\frac{C^{(I,S)}_{\frac{1}{2}^+}}{4\,f^2}\, \Big(
2\,\sqrt{s}-M-\bar M \Big) \,, \label{final-t} \nonumber\\
&& V^{(I,S)}_{0^+}(\sqrt{s}\,) =V^{(I,S)}_{1^+}(\sqrt{s}\,)=
\frac{C^{(I,S)}_{0^+}}{8\,f^2}\, \Big( 3\,s-M^2-\bar M^2-m^2-\bar
m^2
 -\frac{M^2-m^2}{s}\,(\bar
M^2-\bar m^2)\Big) \,, \label{VWT}
\end{eqnarray}
\end{widetext}
where $(m,M)$ and $(\bar m, \bar M)$ are the masses of initial and
final states. We use capital $M$ for the masses of heavy-light
fields and small $m$ for the masses of the Goldstone bosons. The
effective interaction kernel $V^{(I,S)}(\sqrt{s}\,)$ in
(\ref{final-t}) is determined by the leading-order chiral SU(3)
Lagrangian (\ref{WT-term}).

The enormous predictive power of the chiral Lagrangian lies in its specification of the
coupled-channel matrices $C^{(I,S)}_{J^P}$. For the scattering of Goldstone bosons
off any SU(3) anti-triplet field the latter may be classified by invariants according to
the decomposition
\begin{eqnarray}
\bar 3\otimes 8= \bar 3\oplus 6 \oplus \overline{15}\,. \label{}
\end{eqnarray}
The leading order chiral Lagrangian predicts attraction in the anti-triplet and sextet channels but
repulsion for the anti-quindecimet. This holds irrespective whether the target is a meson or baryon. What
matters is that the target transforms as an anti-triplet under the SU(3) flavour group. A further
prediction of chiral SU(3) symmetry is that the attraction in the anti-triplet is thrice as strong as the
attraction in the sextet sector. A similar analysis reveals that the scattering of Goldstone bosons off
sextet targets
\begin{eqnarray}
6 \otimes 8= \bar 3\oplus 6 \oplus \overline{15} \oplus 24 \,,
\end{eqnarray}
is attractive in the anti-triplet, sextet but also in the
anti-quindecimet channels. The interaction is repulsive in the
24-plet channels. Chiral SU(3) symmetry predicts a hierarchy of
strength with strongest attraction in the triplet channels, which
is five  times as strong as the attraction in the anti-quindecimet
channels. In the sextet channel the attraction is reduced by a
factor $3/5$ only. It is also instructive to compare the amount of
attraction in the anti-triplet channels as they result from the
reduction of $\bar 3\otimes 8$ versus $6 \otimes 8$. Chiral
symmetry predicts stronger binding in the latter case. The amount
of attraction is larger by a factor $5/3$ as compared to the
former case \footnote{We correct two minor misprints in Tab. 3 of
\cite{LK04}. The 11 components of the second and fifth row should
read -2 rather than -1 and 0.}.

It is left to specify the loop functions $J^{(I,S)}_{J^P}(\sqrt{s}\,)$ in (\ref{VWT}).
The latter are diagonal in the coupled-channel
space, depend however on whether to scatter Goldstone bosons off
mesons or baryons. For details we refer to \cite{LK02,LK03,LK04,KL04}. The final expressions
are quite transparent,
\begin{widetext}
\begin{eqnarray}
&& J_{0^+}(\sqrt{s}\,) = I(\sqrt{s}\,)-I(\mu_{0^+})\,, \qquad
\qquad
 J_{1^+}(\sqrt{s}\,) = \Big(1 + \frac{p_{\rm }^2}{3\,M^2}
\Big)\, \big[I(\sqrt{s}\,)-I(\mu_{1^+}) \big]\,,
\\
&& J_{\frac{1}{2}^+}(\sqrt{s}\,) = \Big(M + (M^2+p_{\rm }^2)^{1/2}
\Big)\, \big[I(\sqrt{s}\,)-I(\mu_{\frac{1}{2}^+}) \big]\,,
\nonumber\\
&& I(\sqrt{s}\,)=\frac{1}{16\,\pi^2} \left(
\frac{p_{}}{\sqrt{s}}\, \left( \ln
\left(1-\frac{s-2\,p_{}\,\sqrt{s}}{m^2+M^2} \right) -\ln
\left(1-\frac{s+2\,p_{}\sqrt{s}}{m^2+M^2} \right)\right) \right.
 + \left.
\left(\frac{1}{2}\,\frac{m^2+M^2}{m^2-M^2} -\frac{m^2-M^2}{2\,s}
\right) \,\ln \left( \frac{m^2}{M^2}\right) \right)\;, \nonumber
\label{i-def}
\end{eqnarray}
\end{widetext}
where $\sqrt{s}= \sqrt{M^2+p_{}^2}+ \sqrt{m^2+p_{}^2}$. Note
that the loop functions in (\ref{i-def}) specifying the scattering of Goldstone bosons off
$0^-$ and $1^-$ targets differ by a
term suppressed with $1/M^2$ only. A crucial ingredient of the
$\chi-$BS(3) scheme is its approximate crossing symmetry
guaranteed by a proper choice of the subtraction scales,
\begin{eqnarray}
\begin{array}{ll}
\mu_{0^+}^{(I,0)} =  M_{D(1867)}\,, \phantom{xxxxx}&
\mu_{0^+}^{(I, \pm 1)} = M_{D_s(1969)}\,, \\
\mu_{0^+}^{(I, 2)} = M_{D(1867)} \,,&
\mu_{1^+}^{(I,0)} =  M_{D(2008)}\,, \\
\mu_{1^+}^{(I, \pm 1)}= M_{D_s(2110)}\,,&
\mu_{1^+}^{(I, 2)} = M_{D(2008)} \,\\
\mu_{[\bar 3]}^{(I,0)} =  M_{\Lambda_c(2284)}\,,&
\mu_{[\bar 3]}^{(I, \pm 1)} = M_{\Xi_c(2470)}\,,\\
\mu_{[\bar 3]}^{(I, -2)} = M_{\Lambda_c(2284)} \,,&
\mu_{[6]}^{(I,0)} = M_{\Sigma_c(2453)}\,, \\
\mu_{[6]}^{(I,\pm 1)} = M_{\Xi'_c(2580)}\,,&
\mu_{[6]}^{(I, -2)} = M_{\Omega_c(2704)} \,, \\
\mu_{[6]}^{(I, -3)} = M_{\Xi'_c(2580)} \,.&
\end{array} \label{ren-cond}
\end{eqnarray}

\begin{figure}[b]
\begin{center}
\vskip-0.5cm
\includegraphics[width=8.9cm,clip=true]{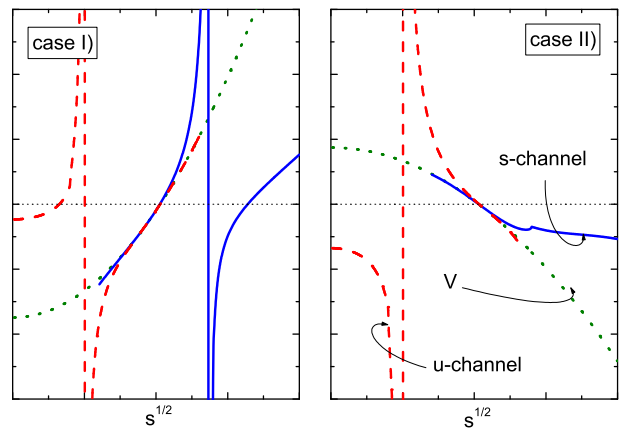}
\end{center}\vspace*{-1.0cm}
\caption{Typical cases of forward scattering amplitudes. The solid
(dashed) line
 shows the s-channel (u-channel)
unitarized scattering amplitude. The dotted lines represent the
amplitude evaluated at tree-level.} \label{fig:crossing}
\end{figure}

The renormalization condition (\ref{ren-cond}) reflects the basic
assumption our effective field theory is built on, namely, that at
subthreshold energies the scattering amplitudes may be evaluated
in standard chiral perturbation theory. Once the available energy
is sufficiently high to permit elastic two-body scattering, a
further typical dimensionless parameter of order one arises, that
is uniquely linked to the presence of a two-particle unitarity
cut. Thus it is sufficient to sum those contributions keeping the
perturbative expansion of all terms that do not develop a
two-particle unitarity cut. In order to recover the perturbative
nature of the subthreshold scattering amplitude the subtraction
scale $ M-m < \mu < M+m$ must be chosen in between the s- and
u-channel elastic unitarity branch points \cite{LK02}.  It was
suggested that s-channel and u-channel unitarized amplitudes
should be glued together at subthreshold kinematics \cite{LK02}. A
smooth result is guaranteed if the full amplitudes match the
interaction kernel $V$ close to the subtraction scale $\mu$ as
implemented by (\ref{ren-cond}). In this case the crossing
symmetry of the interaction kernel, which follows directly from
its perturbative evaluation, is carried over to an approximate
crossing symmetry of the full scattering amplitude. This
construction reflects our basic assumption that diagrams showing
an s-channel or u-channel unitarity cut need to be summed to all
orders typically at energies where the diagrams develop their
imaginary part. In Fig. \ref{fig:crossing} we demonstrate the
quality of the proposed matching procedure as applied for typical
forward scattering amplitudes. The figure clearly illustrates the
smooth matching of s-channel and u-channel iterated amplitudes at
subthreshold energies.

\section{Discussion of results}

We first discuss the resonance spectrum as it arises in the
'heavy' SU(3) limit \cite{Copenhagen,LK03} with $m_{\pi ,K,\eta} =
500$~MeV. Using degenerate masses of $1800$~MeV for the
pseudo-scalar and vector D-mesons an almost degenerate spectrum of
scalar and axial-vector D-mesons is obtained. The mass splitting
are below 1~MeV reflecting the presence of the heavy-quark
symmetry. In this case an anti-triplet of mass 2204~MeV with poles
in the $(I,S)=(0,+1),(1/2,0)$ amplitudes is generated dynamically.
In the sextet channels the attraction is not quite strong enough
to form a bound or clear resonance state signal. However, if the
attraction is increased slightly by using $f=80$~MeV rather than
the canonical value $90$~MeV, poles at mass 2298~MeV arise in the
$(1,+1),(1/2,0),(0,-1)$ amplitudes. Similar results are produced
for the s-wave open-charm baryon resonances. Using a somewhat
arbitrary common mass $2400$~MeV for the anti-triplet ground
states an anti-triplet of mass 2778~MeV with poles in the
$(0,0),(\frac{1}{2},-1)$ amplitudes and a sextet of mass 2900~MeV
with poles in the $(1,0),(\frac{1}{2},-1),(0,-2)$ amplitudes is
generated dynamically. The result is quite stable against small
variations of the optimal subtraction scales of (\ref{ren-cond}).
Lowering the latter by 200~MeV reduces the anti-triplet and sextet
masses by 40~MeV and 5~MeV only. Assuming a common mass for the
sextet ground states of 2500~MeV we obtain an anti-triplet of mass
2807~MeV with poles in the $(0,0),(\frac{1}{2},-1)$ amplitudes, a
sextet of mass 2875~MeV with poles in the
$(1,0),(\frac{1}{2},-1),(0,-2)$ amplitudes and an anti-quindecimet
of mass 3000~MeV with poles in the
$((\frac{1}{2},1),(0,0),(1,0),(\frac{1}{2},-1),(\frac{3}{2},-1),(1,-2))$
amplitudes.

Contrasted results follow in the 'light' SU(3) limit
\cite{Copenhagen,LK03} with $m_{\pi,K,\eta} \sim 140$ MeV. No
clear signal of a resonance or bound state in any of the channels
is found. The striking dependence of the resonance spectrum on the
current quark masses of QCD should be tested in un-quenched
lattice simulations.

\begin{figure}[b]
\begin{center}
\vskip-0.2cm
\includegraphics[clip=true,width=8.5cm]{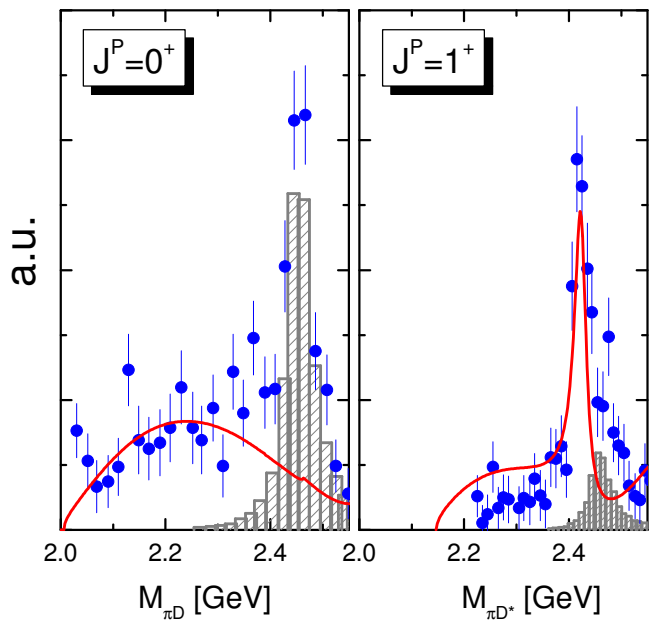}
\vskip-0.6cm
\end{center}
\caption{Mass spectra of the $(\frac{1}{2},0)$-resonances as seen in the
$ \pi \,D(1867)$-channel (l.h. panel $J^P=0^+$) and $ \pi \,D(2008)$-channel (r.h. panel $J^P=1^+$).
The solid lines show the theoretical mass distributions. The data are taken from \cite{BELLE}
as obtained from the $B \to \pi \,D(1867)$  and $B \to  \pi \,D(2008)$
decays. The histograms indicate the contribution from the
$J=2$ resonances $D(2460)$ as given in \cite{BELLE}.}\label{fig:BELLE}
\end{figure}

Realistic results with clear bound-state or resonance signals in
many channels are implied by using physical masses of the ground
state open-charm meson and baryons. We first discuss the
open-charm meson spectrum. In the $(0,1)$-sector we predict a
scalar bound state of mass 2303 MeV and  an axial-vector state of
mass 2440 MeV at leading chiral order. According to
\cite{BEH03,NRZ03} these states should be identified with the
narrow resonances of mass 2317 MeV and 2463 MeV recently observed
by the BABAR collaboration \cite{BaBar,CLEO}. Since we do not
consider isospin violating processes like $\eta \to \pi_0$ the
latter states are true bound state in our present scheme. Given
the fact that our computation is parameter-free this is a
remarkable result.

\begin{figure}[t]
\begin{center}
\vskip-0.2cm
\includegraphics[angle=-90,width=8.5cm,clip=true]{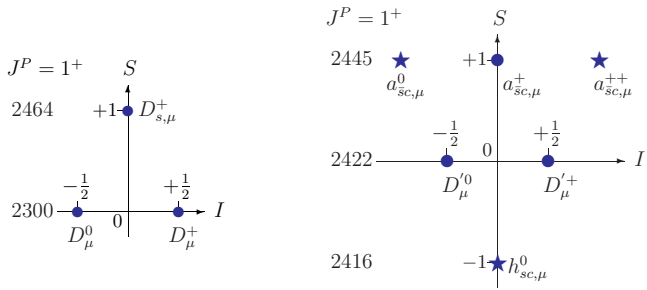}
\vskip-0.4cm
\end{center}
\caption{Isospin ($I$) and strangeness ($S$) quantum numbers of
open-charm mesons with $J^P=1^+$. Masses in units of MeV of the
chiral excitations are indicated. Resonances with exotic quantum
numbers are represented by stars. \label{fig:levels-one-plus}}
\end{figure}

\begin{figure}[t]
\begin{center}
\vskip-0.2cm
\includegraphics[clip=true,width=8cm]{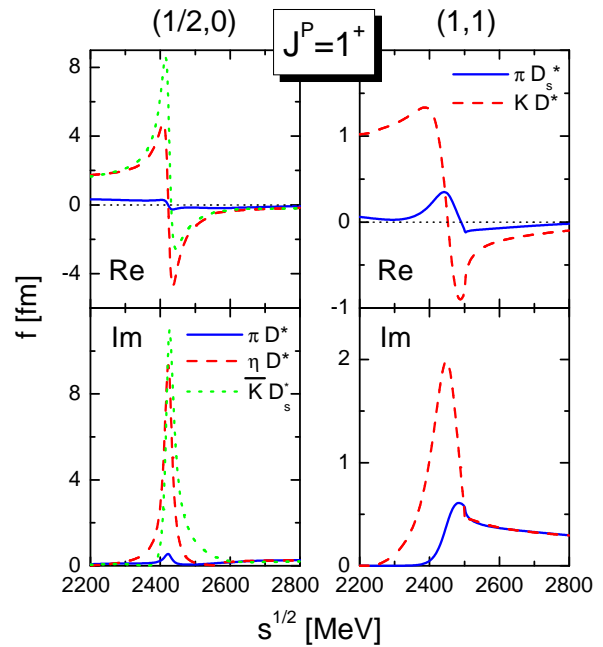}
\vskip-0.6cm
\end{center}
\caption{Open-charm resonances with $J^P =1^+$  and $(I,S)=(\frac{1}{2},0),(0,1)$ as seen in
the scattering of Goldstone bosons of $D(2008)$- and $D_s(2110)$-mesons. Shown are real and imaginary
parts of reduced scattering amplitudes.
}\label{fig:amplitudes-one-plus}
\end{figure}

Further clear scalar and axial-vector resonance signals are seen
in the $(\frac{1}{2},0)$-sector at leading order. Chiral dynamics
predicts a narrow scalar state of mass 2413 MeV just below the
$\eta\,D(1867)$-threshold and a broad scalar state of mass 2138
MeV. Modulo some mixing effects the heavier of the two is part of
the sextet the lighter a member of the anti-triplet. Heavy-quark
symmetry implies an analogous pattern for the axial-vector
resonances. A narrow structure at 2552 MeV arises together with a
broad state at around 2325 MeV. The two broad states, which were
first predicted by the chiral quark model \cite{BR03,NRZ03}, were
confirmed by the recent data of the BELLE collaboration
\cite{BELLE}. In Fig. \ref{fig:BELLE} we display the data together
with the result of \cite{HL04} which performed a chiral
coupled-channel computation at subleading order. In the right-hand
panel three resonances contribute to the measured $\pi\,
D(2008)$-spectrum. The effect of a narrow $2^+$ state of mass 2460
MeV indicated by the histogram is small and not considered in the
computation \cite{HL04}. The remaining two states are identified
with a broad anti-triplet and a narrow sextet state. This implies
that the well established $D(2420)$-resonance \cite{PDG02} should
be a member of a sextet rather than a member of an anti-triplet as
commonly assumed. Of course this questions also the interpretation
of the $2^+$ state of mass 2460  MeV to be the heavy-quark partner
of the $D(2420)$ suggesting at least one additional so far
unobserved state, possibly with quantum numbers $3^+$. To finally
settle this issue requires certainly further work. It should be
stressed that in the computation \cite{HL04} free parameters which
enter at subleading order were adjusted to recover the $D(2420)$
resonance at its physical mass. At leading order the mass of the
latter is overestimated by about 130 MeV even though a small width
of about 20 MeV is obtained. Thus the chiral correction terms
increase the amount of attraction in the sextet channel leading to
the level pattern shown in Fig. \ref{fig:levels-one-plus}.  In
particular a $K D(2008)$-molecule state of mass 2416 MeV, which
carries exotic quantum numbers, is predicted. We suggest to call
the latter $h_{sc}$ in analogy to the axial-vector isospin-singlet
state $h(1170)$. Similarly, the name $a_{\bar s c}$ for the
isospin-triplet members of the sextet is proposed. The various
reduced scattering amplitudes are shown in channels where the
generated states have a finite width are shown in Fig.
\ref{fig:amplitudes-one-plus}. The figure clearly illustrates the
typical phenomenon observed in chiral coupled-channel theory: the
$(I,S)=(1/2,0)$-state couples most strongly to the kinematically
closed channels $\eta D(2008)$ and $\bar K D_s(2110)$.

\begin{figure}[b]
\begin{center}
\vskip-0.4cm
\includegraphics[angle=-90,width=8.4cm,clip=true]{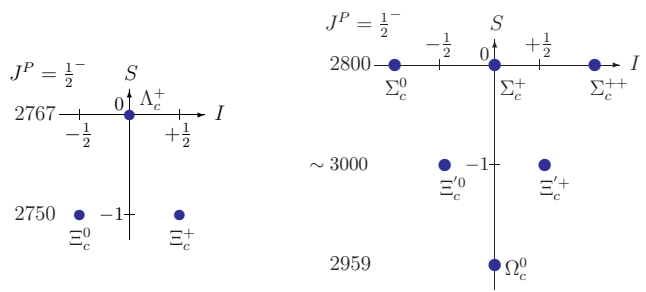}
\vskip-0.5cm
\end{center}
\caption{Isospin ($I$) and strangeness ($S$) quantum numbers of
open-charm baryons with $J^P=\frac{1}{2}^-$ generated dynamically
by scattering Goldstone bosons off the triplet ground states.
Masses in units of MeV of the chiral excitations are indicated.
\label{fig:anti-triplet-baryons}}
\end{figure}

A spectrum similar to the one for the axial-vector mesons arises
for the open-charm scalar mesons. At leading order two
isospin-doublet resonances with zero strangeness are generated by
coupled-channel effects. The triplet state couples strongly to the
$\pi D(1868)$-channel leading to its large width. The sextet state
couples only weakly to the latter channel implying a narrow state.
In Fig. \ref{fig:BELLE} an extreme scenario is shown in which the
sextet state decouples from the $\pi D(1868)$-channel all
together. This can be achieved upon considering chiral correction
terms \cite{HL04}. The sextet state with zero strangeness sits at
2389 MeV in this case. The $K D(1867)$ molecule state comes at
2352 MeV. The signal of the sextet is weakest in the $(I,S)=(1,1)$
sector where the scattering amplitude shows a strong cusp effect
close to the $K D(1867)$-threshold only. We emphasize that it is
difficult to make a precise prediction for the width of the
iso-scalar sextet state. If we only slightly change the set of
parameters the sextet state shows up as a narrow peak in the $\pi
D(1867)$-spectrum of Fig. \ref{fig:BELLE}. Depending on the
precise values of the parameters this state may be detected most
efficiently via its coupling to the $\eta \,D(1867)$ channel
utilizing the $\eta - \pi_0$-mixing effect. In any case it is of
utmost importance to further improve the quality of the empirical
spectrum. The existence of scalar open-charm mesons with exotic
quantum numbers was suggested first by Terasaki based on a
phenomenological constituent diquark approach \cite{Terasaki}.

\begin{figure}[b]
\begin{center}
\vskip-0.2cm
\includegraphics[clip=true,width=8.3cm]{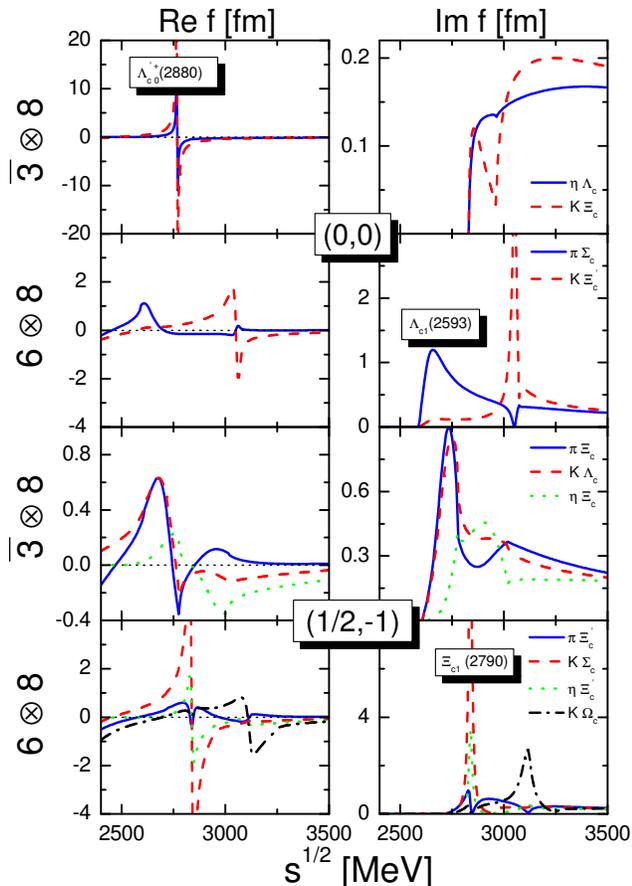}
\vskip-0.8cm
\end{center}
\caption{Open-charm baryon resonances with $J^P=\frac{1}{2}^-$  and $(I,S)=(0,0)$ and
$(1/2,-1)$ as seen in
the scattering of Goldstone bosons off anti-triplet $(\Lambda_c(2284), \Xi_c(2470))$ and
sextet $(\Sigma_c(2453),\Xi'_c(2580),\Omega_c(2704))$ baryons.
Shown are real and imaginary parts of reduced scattering amplitudes.}
\label{fig:triplet}
\end{figure}

\begin{figure}[b]
\begin{center}
\vskip-0.2cm
\includegraphics[clip=true,width=8.3cm]{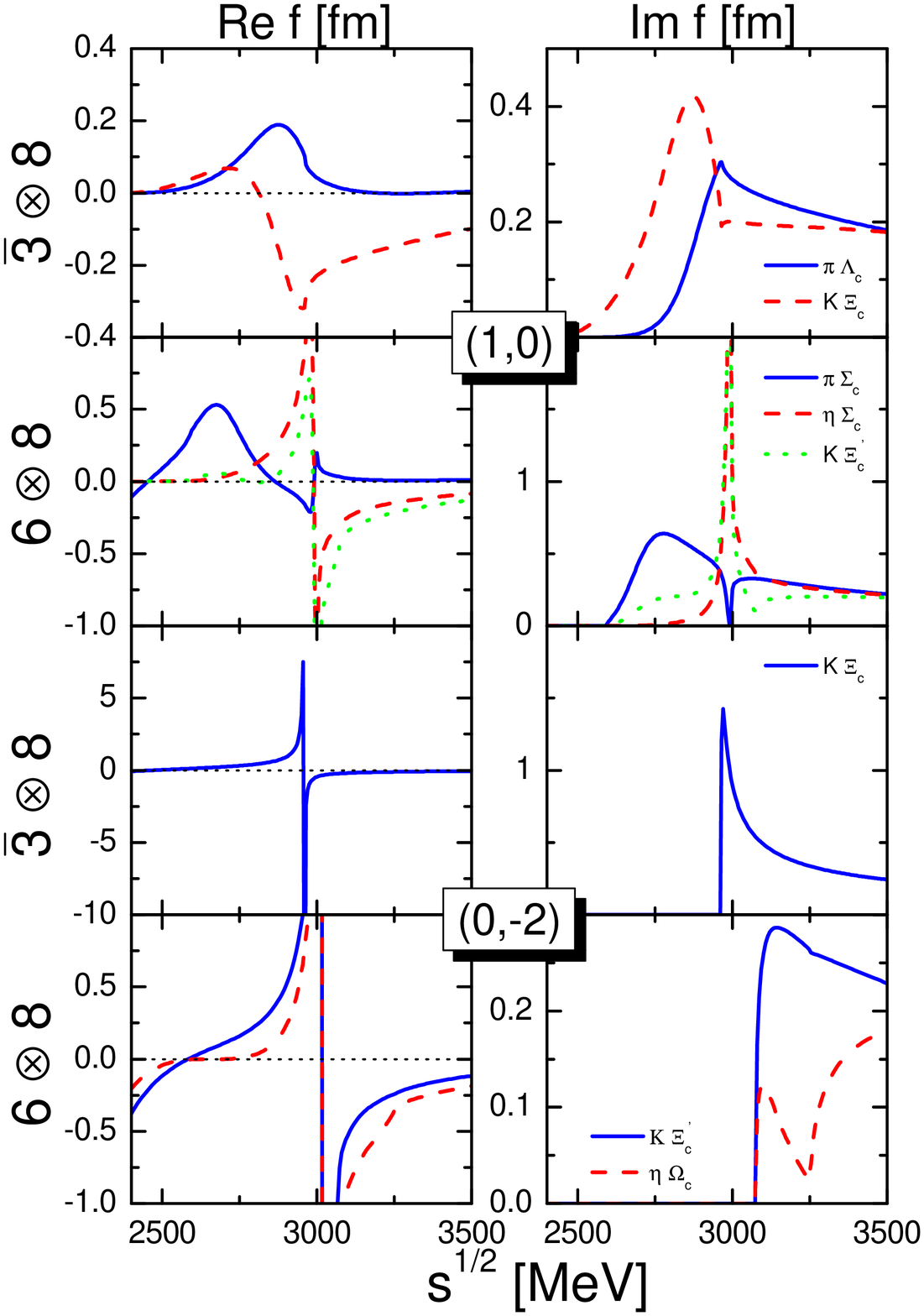}
\vskip-0.8cm
\end{center}
\caption{Open-charm baryon resonances with $J^P=\frac{1}{2}^-$ and
$(I,S)=(1,0)$ and $(0,-2)$ as seen in the scattering of Goldstone
bosons off anti-triplet $(\Lambda_c(2284), \Xi_c(2470))$ and
sextet $(\Sigma_c(2453),\Xi'_c(2580),\Omega_c(2704))$ baryons.
Shown are real and imaginary parts of reduced scattering
amplitudes.} \label{fig:sextet}
\end{figure}

\begin{figure*}[t]
\begin{center}
\vskip-0.2cm
\includegraphics[angle=-90,width=16.5cm,clip=true]{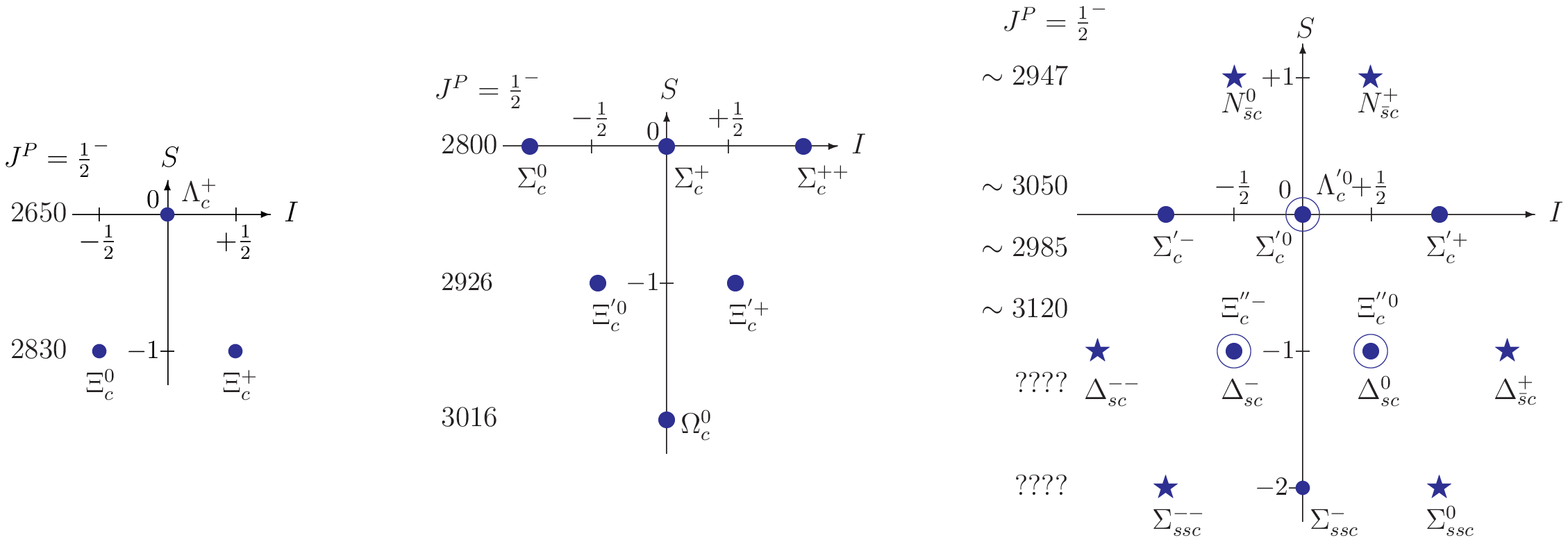}
\vskip-0.6cm
\end{center}
\caption{Isospin ($I$) and strangeness ($S$) quantum numbers of
open-charm baryons with $J^P=\frac{1}{2}^-$ generated dynamically
by scattering Goldstone bosons off the sextet ground states.
Masses in units of MeV of the chiral excitations are indicated.
Resonances with exotic quantum numbers are represented by
stars.\label{fig:sextet-baryons}}
\end{figure*}

We turn to the chiral excitations of open-charm baryons.
Unfortunately, at present there is very little known empirically
about the open-charm baryon resonance spectrum. Only three states
$\Lambda_{c1}(2593), \Lambda_{c}(2880)$ and $\Xi_{c}(2790)$ were
discovered by the CLEO collaboration so far \cite{PDG02}. The
level scheme, as it is predicted by chiral dynamics at leading
order using physical values for the ground state masses, is drawn
in Fig. \ref{fig:anti-triplet-baryons} for the s-wave resonances
generated by the scattering of Goldstone bosons off the
anti-triplet ground states. We obtain a bound state of mass 2767
MeV and quantum numbers $(I,S)=(0,0)$. This state should be
identified with the $\Lambda_c(2880)$ recently detected by the
CLEO collaboration \cite{Artuso} via its decay into the $\pi
\Sigma_c(2453)\to \Lambda_c\pi\pi$ channel. The narrow width of
the observed state of smaller than 8 MeV \cite{Artuso} appears
consistent with a suppressed coupling of that state to the $\pi
\Sigma_c(2453)$ channel as predicted by chiral symmetry. The
various scattering amplitudes in the $(0,0)$ and $(1/2,-1)$
sectors are collected in Fig. \ref{fig:triplet}. Similarly the
scattering amplitudes \footnote{The second row corrects a slightly
incorrect figure shown in \cite{LK04}.} probing the sextet states
of Fig. \ref{fig:anti-triplet-baryons} are collected in Fig.
\ref{fig:sextet}. The figures clearly indicate that it is much too
simple showing a schematic level diagram only. The amplitudes
exhibit a complicated multi-channel structure. For instance
whereas the scattering amplitudes show a clear signal for the
SU(3) triplet state $\Xi_c(2750)$ a corresponding state belonging
to the sextet is quite broad and does not manifest itself very
clearly in the amplitudes of Fig. \ref{fig:triplet}. The existence
of the sextet may be best confirmed searching for an
isospin-singlet $K\,\Xi_c(2470)$-molecule.

\begin{figure}[b]
\begin{center}
\vskip-0.3cm
\includegraphics[clip=true,width=8.5cm]{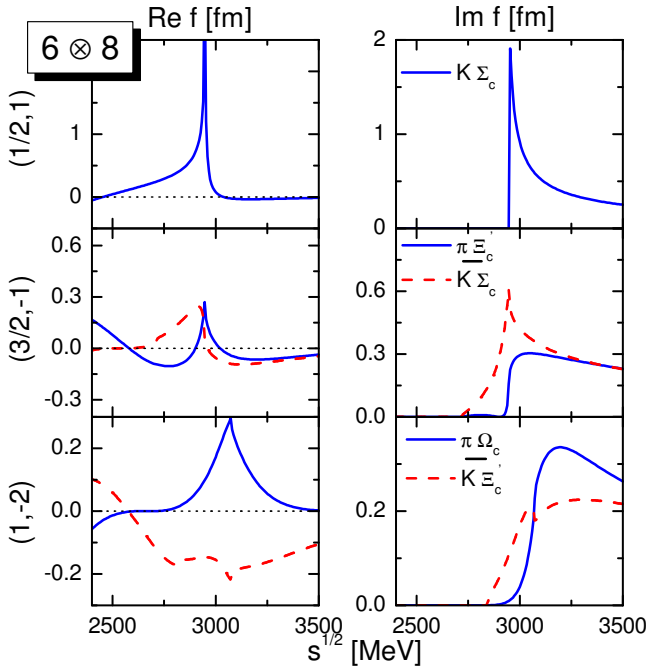}
\vskip-0.7cm
\end{center}
\caption{Open-charm baryon resonances with
$(I,S)=(\frac{1}{2},1),(\frac{3}{2},-1),(1,-2)$ and
$J^P=\frac{1}{2}^+$  as seen in the scattering of Goldstone bosons
off sextet $(\Sigma_c(2453),\Xi_c(2580),\Omega_c(2704))$ baryons.
Shown are real and imaginary parts of reduced scattering
amplitudes.} \label{fig:15plet}
\end{figure}

We conclude our discussion with the spectrum generated dynamically
by scattering Goldstone bosons off the sextet ground states. The
resulting level scheme is drawn schematically in Fig.
\ref{fig:sextet-baryons}. Since the interaction is attractive in
the anti-triplet, sextet and anti-quindecimet channels a quite
rich spectrum arises. The figure suggests a particular labeling of
the exotic multiplets. In particular we denote an isospin-triplet
state with strangeness minus-two by $\Sigma_{s s c}$. Similarly we
introduce the isospin-quartet, $\Delta_{s c}$, which carries
strangeness minus-one. Again the level scheme has to be taken with
great care since not in all multiplet channels clear resonance
structures are expected. It is an important part of the predictive
power of chiral coupled-channel dynamics to spell out in which
channels one expects clear signals. Full details of the resonance
properties are provided by Figs.
\ref{fig:triplet}-\ref{fig:15plet} in terms of reduced scattering
amplitudes. For the channels in which the chiral excitations of
the anti-triplet and sextet manifest themselves with states of
identical isospin and strangeness quantum numbers the
corresponding amplitudes are grouped together (see Figs.
\ref{fig:triplet}-\ref{fig:sextet}). This is instructive since at
subleading order the amplitudes start to couple.

Consider first the anti-triplet states that are generated
dynamically as chiral excitations of the sextet ground states.
Since the attraction is largest in this channel one expects
strongly bound systems with $(I,S)=(0,0)$ and $(1/2,-1)$. Indeed
the second row of Fig. \ref{fig:triplet} shows a resonance at
about 2650 MeV that couples strongly to the $\pi \Sigma_c(2453)$
channel. The properties of this state are close to the ones of the
$\Lambda_c(2593)$ resonance \cite{Edwards}. Here we obtain a decay
width which is significantly larger than the empirical width of
about 4 MeV \cite{Edwards}. Chiral correction terms that couple
the states seen in the  1st and 2nd rows of Fig. \ref{fig:triplet}
are expected to decrease this width. Level-level repulsion of the
two observed states should lower the mass of the lighter state but
push up the mass of the heavier state. In the 4th row of Fig.
\ref{fig:triplet} a narrow state at 2830 MeV is shown. This state
couples strongly to the $K\,\Sigma_c$ and $\eta \,\Xi_c$ channels
and due to its small width should be identified with the
$\Xi_c(2790)$ resonance \cite{Csorna}.

The sextet manifests itself with a bound-state of mass 3015 MeV with $(I,S)=(0,-2)$ as
illustrated in the 4th row of Fig. \ref{fig:sextet}. This channel is a unique probe
for the sextet since neither the ant-triplet nor the anti-quindecimet contributes here.
A further broad sextet state of mass $\sim$2800 MeV with $(1,0)$ is seen in the 2nd row
of Fig. \ref{fig:sextet}. It couples strongly to the $\pi \Sigma_c(2453)$-channel.
Finally the second broad resonance of mass 2926~MeV in the 4th row of Fig. \ref{fig:triplet}
that couples strongly to the $\pi \Xi_c(2580)$-channel
should also be a member of the sextet.

The most exciting consequence of chiral coupled-channel dynamics
is its prediction of anti-quindecimet states. As indicated in Fig.
\ref{fig:sextet-baryons} we do not expect clear signals in all
channels. Most prominent are the states with $(I,S)=(0,0),
(1,0),(1/2,-1)$ around 3 GeV as shown in the 2nd rows of Figs.
\ref{fig:triplet}-\ref{fig:sextet} and 4th row of Fig.
\ref{fig:triplet}. Unfortunately these states do not carry exotic
quantum numbers. The channels in which the anti-quindecimet
manifests itself with exotic quantum numbers are displayed in Fig.
\ref{fig:15plet}. One may hope that chiral corrections terms
conspire to further increase the amount of attraction leading to
signals that are easier to observe.

\section{Summary}

We reviewed the spectrum of chiral excitations in the open-charm meson and baryon sector.
Exciting new predictions for the existence of exotic multiplets were presented in detail.
These results suggest a dedicated search for such states but also require extensive
chiral coupled-channel computations to further substantiate the theoretical predictions.


\end{document}